\documentclass{ws-p8-50x6-00}

\newcommand{\beq}{\begin{equation}}
\newcommand{\eeq}{\end{equation}}

\begin{document}

\title{Kolmogorov Complexity, Cosmic Background Radiation and Irreversibility}

\author{V.G.Gurzadyan}

\address{Yerevan Physics Institute and Garni Space Astronomy Institute, Armenia\\
ICRA, Department of Physics, University of Rome La Sapienza, Rome, Italy}

\maketitle

\abstracts{We discuss the algorithmic information approach to the analysis of
the observational data on the Universe. Kolmogorov
complexity is proposed as a descriptor of the Cosmic Microwave Background (CMB)
radiation maps. An algorithm of computation of the complexity is described,
applied, first, to toy models and then, to the data of the Boomerang experiment.
The sky maps obtained via the summing of two independent Boomerang channels reveal
threshold independent behavior of the mean ellipticity of the anisotropies,
thus indicating correlations present in the sky signal and possibly carrying
crucial information on the curvature and the non-Friedmannian, i.e.
accelerated expansion of the Universe. Similar
effect has been detected for COBE-DMR 4 year maps. 
Finally, as another application of these concepts, 
we consider the possible link between the CMB properties, curvature of the Universe
and arrows of time.}

\vspace{0.1in}
{\it Talk at XXII Solvay Conference on Physics "The Physics of Communication" (Delphi, November 24-29, 2001), in Proceed. Eds. I.Antoniou, V.A.Sadovnichy, H.Walter, World Scientific, pp.204-218, 2003.}

\section{Introduction}

In the process of the study of the Nature we are dealing not with
photons, atoms, planets, galaxies and so on, but with the information that
we are obtaining on them. If one can succeed to get a 'complete' information
(in bits) on a given object, then, in principle, one can recover
that object. Obviously, the main problem is whether
one can define a universal device, so that the coding and decoding of
the information will be independent on the device. 
The complexity or algorithmic information \cite{kolm}\cite{chaitin} 
carries this idea,
i.e. invariant description of an object with respect to a universal computer.
The existence of a universal computer was proven by Kolmogorov.

Invariant descriptors have always been of outstanding importance in any research. 
One can recall the discovered by De Moivre, Laplace and Gauss independence of
the behavior of errors on the phenomena, among the latter,
the coin tossing, human birth rates and observations of planetary motions.

In the present article we apply the concepts of algorithmic information
theory in the studies of the properties of the Universe, via
the analysis of the Cosmic Microwave Background (CMB) radiation data.

CMB provides one of the key windows to the early history of the Universe. 
In accordance with the currently held views, the expanding and cooling
Universe at certain moment of time after the Big Bang at redshifts $z\simeq 1100$, becomes 
transparent for photons which we detect now, at $z=0$, as CMB with highly isotropic temperature 
and accurate Planckian spectrum. 
The discovery of the quadruple anisotropy of the CMB by Cosmic
Background Explorer (COBE) satellite in 1992 marked an important phase
for cosmological studies.
The next generation of experiments, Boomerang, Maxima, DASI, CPI,
measured the peaks of the power spectrum (angular autocorrelation function). Existence of
such peaks was predicted decades ago as a consequence
of the interplay of the periods of the compression waves and the decoupling time
at the epoch of the last scattering. 
On the one hand, this supports the basic paradigms of the Big Bang cosmology,
on the other hand, the existence of over 10 free parameters shows the necessity of
extreme care in the determination of unambiguous values of the key cosmological parameters.     
Among the latter parameters which are evaluated only in combination with others,
is the curvature of the Universe. Such degeneracy makes
hard to get an empirical prove of the precise flatness of the Universe.
 
CMB sky maps, i.e. the distribution of the temperature anisotropies over the sky, have been studied via various descriptors mainly to trace possible departures of the noise from the Gaussian one.
The use of Kolmogorov complexity to extract cosmological information from the CMB
maps has been proposed in \cite{gur}. It was motivated by the possibility of
tracing of the geometry of the Universe by means of the effect of geodesic mixing.
The effect of geodesic mixing, which is absent at precisely flat and positively curved spaces, can therefore open a way to distinguish the curvature. 

Our broader aim here however, is to illustrate
the efficiency and universality of complexity in the analysis
of profound astrophysical problems. We complete our discussion
with consideration of even 
older problem, the irreversibility and arrows of time.

We start with a brief account of the effect of geodesic mixing, 
definition and properties of complexity and then move to the description
of algorithm of computations for CMB maps, with  subsequent
application to the Boomerang data \cite{bern}.

\section{Geodesics}

Let us first inquire, when the geodesic defined on
a $4$--dimensional manifold $W$ with Lorentzian metric $^4{\bf g}$ 
which is oriented and time--oriented, will be a
geodesic of a $3$--dimensional manifold $M$ of Riemannian metric $^3 \bf h$
with respect to the operation of projection.

The projection $\pi$ is defined uniquely as
$$
\begin{array}{cc}
   \begin{array}{rcl}
    W & \stackrel{\textstyle \jmath^{-1}}{\rightarrow}& M\times R\\
      &                          {\searrow}           & \downarrow\pi_M\\
      & \stackrel{\textstyle\pi}{}                    & M
   \end{array} & \qquad \pi=\pi_M\circ\jmath^{-1},
\end{array}
$$
where
$$
    \pi_M:M\times R\to M:(x,t)\mapsto x.
$$
This reduces the curve $\gamma$ in $W$ to a
curve $c$ in $U\in M$ as represented below
$$
\begin{array}{cc}
   \begin{array}{rcl}
                W & \stackrel{\textstyle \pi}{\rightarrow}   & M\\
  \gamma \uparrow & \stackrel{\searrow}{\textstyle\phi}   & \uparrow c\\
                R & \stackrel{\rightarrow}{\textstyle\sigma} & R
      \end{array} 
  &\qquad 
          \begin{array}{ll}
             &\sigma=\phi\circ\gamma:R\to R \mbox{ is a diffeomorphism}\\
             &c=\pi\circ\gamma\circ\sigma^{-1}:R\to 
M:\lambda\mapsto c(\lambda).
          \end{array}
\end{array}
$$ 
The search of the conditions to be satisfied in order the projection of 
any null geodesic on $W$ to be a geodesic on $M$ in general case is a complicated problem.
For homogeneous-isotropic spaces this can be traced via the problem of ``internal time"
considered initially in \cite{LMP} where the corresponding sufficient conditions have been obtained. 
For any null geodesics on $W$ to be a geodesic on $M$ the conditions include 
\begin{eqnarray}
&&{\bf g}=a^2(t)\cdot{\bf h},\\
&&\phi(x,t)=\phi_0+\int_{t_0}^t a^{-1}(s)ds,\\
&&\lambda(t)=\int_{t_0}^t a^{-1}(s)ds=\eta(t)-\eta(t_0),
\end{eqnarray}
where $^3{\bf h}$ is one of the metrics of maximally symmetric 
$3$--dimensional manifold. 

Thus the geodesic flow $W=U\times R$ be ($3+1$)--dimensional manifold with  
Robertson--Walker metric can be reduced to a geodesic 
flow on three--dimensional closed manifold $M=U$ 
with metric $a_0^2{\bf h}$ and affine parameter $\lambda$ (for details see \cite{book}).

If the Friedmann-Robertson-Walker universe has a negative curvature,
the flow of null geodesics which describes the free motion of
photons, represents an Anosov system \cite{anosov} (locally, if $U$ is not compact), a class of dynamical systems
with maximally strong statistical properties. Anosov systems are characterized
by exponential divergence of initially close trajectories,
positive KS-entropy, countable Lebesgue spectrum, exponential decay
of time correlation functions and K-mixing.

One of the significant properties of Anosov systems
is the structural stability (coarseness) as proved by Anosov in 1967, 
roughly, the robustness of properties with
respect to perturbations. This is a crucial property also in our context, 
since we live not in a FRW Universe
with strongly constant curvature but with small perturbations of metric, and 
moreover, we know  the magnitude of their smallness from the same CMB measurements.  

  The deviation of two geodesics in $3$--manifold implies that
\begin{equation}
    l(\lambda)=l(0)\exp(\chi\lambda),
\end{equation}
where Lyapunov exponent $\chi=a_0^{-1}$ and $\chi=0$ when $k=0$ or $k=+1$.
Hence 
$$
  \lambda_c^l=\frac{1}{\chi}.
$$
For the geodesic flows in $W$ we have 
\begin{equation}
   L(t)=L(t_0)\frac{a(t)}{a(t_0)}\exp(\chi\lambda(t)).
\end{equation}       
The time correlation function of those geodesic flows  for $d=3$ 
$$
   b_{A_1,A_2}({\lambda})=\int_{SM}A_1\circ f^{\lambda}\cdot A_2 d\mu
         -\int_{SM}A_1d\mu\int_{SM}A_2d\mu,
$$
decreases by exponential law, i.e., $\exists c> 0$ such that for all 
$A_1,A_2\in L^2(SM)$
but a finite-dimensional space in $L^2(SM)$ \cite{po}
\begin{equation}
   \left|b_{A_1,A_2}({\lambda})\right|\leq c\|A_1\|\cdot\|A_2\|\cdot e^{-h{\lambda}} \ ,
\end{equation}
where $d\mu$ is the Liouville measure,
$h$ is the KS-entropy of the geodesic flow $\{f^{\lambda}\}$, and
$$
  \|A\|=\left[\int_{SM}A^2d\mu\right]^{1/2}\ .
$$
One may readily see that if 
$$
  A_1(u)=T(u)\ ,
$$
and
$$
  A_2(u)=\chi_{{\cal K}(v)}(u) \ ,
$$
where $T$ is the temperature of sky at $u$, ${\cal K}(v)$ is a  Cartesian product
of $3D$ ball and $2D$ rigid angle at the point $v$, 
and  $\chi_{{\cal K}(v)}$ is the characteristic function of the set ${\cal K}(v)$,
then
\begin{equation}
  T_{\lambda}(u)=\frac{1}{\mu({\cal K}(u))}\int_{{\cal K}(u)}T\circ f^{\lambda}d\mu
\end{equation}
and
$$
  T_{\lambda}(u)-\bar{T}=\frac{1}{\mu({\cal K}(u))}\cdot b_{A_1,A_2}({\lambda})\ ,
$$
where
$$
  \bar{T}=\int_{SM}Td\mu\ .
$$
Therefore
\begin{equation}
  \left|\frac{T_{\lambda}(u)}{\bar{T}}-1\right|
         \leq\frac{c}{[\mu({\cal K}(u))]^{1/2}}\cdot e^{-h{\lambda}} \ .
\end{equation}
In particular, for any $u$ we have
$$
  \lim_{\lambda\to\infty}T_{\lambda}(u)=\bar{T}\ ,
$$
i.e. the isotropic state is the final state.

Thus the non-zero negative curvature will lead to the decrease of perturbations of the geodesic flows, i.e. of the amplitude of anisotropy 
of CMB, while the strong statistical properties of Anosov systems will
lead to the complexity of the anisotropy areas \cite{gk}.

\section{Kolmogorov Complexity}

A crucial concept for definition of the complexity is that of the
universal computer.
A computer is considered {\it universal}, if for any computer $C$
there exists a constant $S_C$ which can be added to any program $p$,
so that $S_Cp$ should execute the same operation on computer $U$ as
the program $p$ on computer $C$.
The computer is a device performing only deterministic operations,
so that the Turing machine can be considered as an example of universal computer, 
as well as the probabilistic computers of Shennon, which
are using {\it the random rules} to reduce the time of computation
for problems with unique solution. 

The {\it algorithm} is the set of instructions defining
which operations have to be executed by the computer and when.
Since the computer must halt,  the program  cannot be a prefix
for some other program;
a word $a$ is called prefix for a word $b$ if $b=ac$ with some
other word $c$. Hence, the set of accessible programs should be
{\it prefix-free}.

The {\it complexity} $K_U(x)$ of the sequence $x$ with respect to a
universal computer $U$ is defined as the length in bits of the
smallest algorithm $p$ by which the computer $U$ starting with some
{\it initial
fixed state} calculates the object $x$ as its {\it only output}, and
{\it halts}. The complexity was introduced by Kolmogorov, Solomonoff and Chaitin
\cite{kolm},\cite{chaitin}.
The sequence is called complex if its complexity is comparable
with its length.
Note, that the time of calculation is not entering this definition.

The complexity is related with another
basic concept, {\it random sequences}. The most general definition by
Martin-L\"{o}f \cite{martin} is formalizing the idea of Kolmogorov
that random
sequences have very small number of rules comparing to its length; the rule
is defined as an
algorithmically testable and rare property of a sequence. Though correlated
for typical objects, the properties of complexity and randomness
are not identical, however.

Chaotic system which are non-compressible, therefore possess higher complexity
than a regular one which are compressible. As shown by Martin-L\"{o}f, the complexity of
finite sequences varies between $N$ and $N-log_2N$, since even random sequences
can have extended non-random subsets. In such cases the
{\it specific complexity} introduced by Alekseev
\begin{equation}
k(A)=\frac{K_A}{|A|},
\end{equation}
enables to distinguish the random sequences and hence algorithmically complex systems, 
i.e. when at large N a finite limit does exist 
$$
k(N)\rightarrow k\not= 0,
$$ 
from the non-random sequences when this limit is zero.
Random sequences are indistinguishable (for all practical purposes) from
the ones generated by the proper stochastic process \cite{zl}.

In certain trivial cases low-complexity objects can be distinguished
easily, for example, (0,...,0) or (1,...,1).  In some other cases, the
object could have a complex binary representation, such as
$\pi$, though actually being again of low-complexity.
In general case,
however, the situation is much less simple. Moreover, it is proved that
there is no a short algorithm to decide whether a given complex-looking
sequence is really complex \cite{chaitin},\cite{zl}.
Fortunately, though in general the shortest program cannot be reached,
i.e. the exact complexity cannot be calculated, in certain problems
the obtained results cannot be too far from that value.

If the length of a sequence $x$ is $N$ then the obvious upper
limit can be established
\footnote{Note that, if $x$ is the binary representation
of some integer $N_0$, then $N\approx \log _2N_0$.}
\begin{equation}
\label{1}
K_U(x)<N.
\end{equation}
Let us estimate the fraction of such sequences among all
N-bit sequences, for which
$$
K_U(x) < N-m.
$$
This means that there exists a program of length $N-m$ which computes $x$.
The total number of programs of such a length cannot be larger than
$2^{N-m+1}$; this is the upper limit without taking into account the
prefix-free condition. Thus, we have the following upper limit
$$
(2^{N-m+1}-1)/ 2^{-N}\approx 2^{-m+1}.
$$
This value is small if $m$ is sufficiently large. Thus a more general
relation than (\ref{1}) can be established
\begin{equation}
K_U(x)\approx c(x)\, N, c(x)\approx 1
\end{equation}
Thus, the calculation of relative complexity of an object and of a
perturbed object via given computer and developed code (though the latter
cannot be proved to be the shortest possible), has to reflect the
complexity introduced by the perturbation. Since in our problem the
complexity is a
result of propagation of photons after the last scattering surface (if
k=-1), one can thus 'measure the perturbation' caused by the curvature
of the space as it was performed while measuring the elongation of
the CMB anisotropy areas in \cite{gt}.

The complexity of a dynamical system can be determined
by means of the representation of the trajectory
via a symbolic language \cite{arnold}.
Then a trajectory of the considered dynamical system can be viewed as a
sequence of symbols which can be
translated into the language of bits. The dynamics
can be called
chaotic (for fixed initial conditions) if the corresponding
symbolic sequence is algorithmically complex. 
Note that the partition should
be detailed enough because algorithmic complexity is well-defined only for
sufficiently long sequences of symbols.
The CMB digitized maps when given values of averaged temperature are assigned to the pixels 
covering a region of sky is a proper example for symbolic dynamics, and hence
can be linked not only with complexity but also random sequences.

Below we describe an algorithm of estimation of
the complexity for anisotropy areas of computer-imitated CMB maps.
Similarly, one can formulate the problem for definition and
study of the random sequences of CMB maps.

\section{The Complexity Algorithm}

Strictly speaking we can estimate only the upper limit of $K$ corresponding
to a given algorithm. By algorithm we will understand the computer
program in PASCAL \cite{ags}, along with the data file,  describing the coordinates of
the pixel of the anisotropy area (spot).  Namely the data file includes
compressed information about the string of digits.
The program is a sequence of commands performing reconstruction of the string
and calculations of the corresponding lengths. Since at the analysis of
various areas we use the same code, the only
change will be in the data files. Hence the complexity of the figure will
be attributed to the file containing the information on the position of
the pixels.

The code describing the area works as follows. As an initial pixel we fix
the upper left pixel of the area and move clockwise along its boundary.
Each step  -- a 'local step' --  is a movement from a current pixel to the
next one in above given direction. This procedure is rigorously defining the
'previous' and 'next' pixels. Two cases are possible. First, when
the next pixel (or several pixel areas) after the initial one is in the same
row: we write down the number of pixels in such 'horizontal step'.
The second case is, when the next pixel is in vertical direction;
then we perform the local steps in vertical direction ('vertical step') and
record the number of corresponding pixels. Via a sequence of
horizontal and vertical steps we, obviously, return to the initial pixel,
thus defining the entire figure via a resulting data file.

The length of the horizontal step cannot exceed the number of
columns, i.e. $N$, while the vertical step cannot exceed $M$, requiring
$log_2M$ and $log_2N$ bits of information, correspondingly.
For the configurations we are interested in, the lengths of the horizontal
and vertical steps, however,
are much less than $log_2M$ and $log_2N$ and therefore we need a
convenient code for definition of the length of those steps. The code
realized in \cite{ags} was for $M=N=256$; apparently for each value 
of $M$ and $N$ one has to choose the most efficient code.

Thus, after each step, either horizontal or vertical, certain amount of
bits of information is stored. The first two bits will contain information
on the following bits defining the length of the given step in a manner
given in Table 1.
\begin{table*}
\centering
\caption{}
\medskip
\begin{tabular}{lll}
\hline
\hline
first 2 bits & next bits   & \\
\hline
0 1          & 1           & \\
1 0          & 2           & \\
1 1          & 3           & \\
\hline
\end{tabular}
\end{table*}
The case when the first two bits are zero, denotes: if the following digit
is zero than the length of the step is $l_s=0$, and hence no digits of the
same step do exist; if the next digit is $1$, than 8 bits are following, thus
defining the length of the step. If $l_s=1$, than after the combination
$0\, 1$ the following digit will be either $0$ or $1$ depending
whether the step
is continued to the left or to the right with respect to the direction of
the previous step. 

Thus, the complexity is a calculable quantity for CMB digitized sky maps
\cite{ags}.  Its values
correlate also with the values of the fractal dimension of the areas.  

\section{Cosmic Background Maps}

The available CMB maps such as of COBE and even Boomerang, are not enough accurate for
the meaningful calculation of the Kolmogorov complexity, however, its simplified descriptor
can be used for those maps, namely, the mean elongation-ellipticity of the
anisotropy spots.  This aim also needs the development of special algorithms
and careful runs to distinguish correlations from the foreground effects. 

We now briefly describe the special purpose adaptive software MAP08 \cite{kashin}, which
was used for the analysis of the Boomerang data. The code enabled to reveal the hot and cold anisotropy areas, determine their coordinates, sizes in pixel numbers, analysis of their shapes, of the spatial correlation functions, etc.  The software runs in interactive regime with two input datasets A and B obtained at measurements at two channels, via the sum and difference maps, as well as Gaussian and any simulated map.  

After the definition of the temperature threshold interval,
for example, from -2000 $\mu K$ up to +2000 $\mu K$, with step 25 $\mu K$, 
the choice of the minimal and
maximal number of pixels per area (e.g. 3-200), of the step of the input matrix (e.g. 1 arc min),
the mode SkyMap enables the visualization of the map and a creation of a matrix MM containing the pixel data for positive (negative) thresholds of areas of equal or higher (lower) the given threshold.
For parameters of the Boomerang data, the following choice was efficient: 1 Pix – $2 \times 2$ pixels, Sqw - a square of $8' \times 8'$ (approximately), Dia - a circle of diameter 11.2'.
        
The matrix MM [x, y] of a unit step and given size, e.g. $1692 \times 1296$ cells for Boomerang data of $1'$ cell size, defined by the following formula
$$
	X = round (60.0 * (XY [n, 1]-237.11))
$$
$$
  Y = round (60.0 * (XY [n, 2] +41.61));
  MM [x, y] =1,
$$
is determining the anisotropy areas via the choice of pixels with temperatures equal and higher/lower of the given threshold. The non-equal size of the pixels and some other input inhomogeneities are taken into account here, while using both Cartesian or curvilinear coordinates.
The scheme of the regularization can be performed for each coordinate frame, with the subsequent check of its efficiency by means of the least number of the abandoned (non-regularized) pixels.
              
The matrix MM is then scanned, e.g. a field of $\pm11$ pixels in vertical and horizontal directions at a step $1'$, and $\pm1$ cell with a step $7.5'$ (in both cases within a field $22' \times 22'$) with the center in $x_0$ and $y_0$. The points of the array with a given code and current coordinates $x_t$ and $y_t$ are chosen and the determination of the center and other characteristics of the revealed (anisotropy) areas can be performed readily. For example, the condition 
$r_t < = rw$ has to be checked for its each pixel, where $r_t$ is the distance between the current point and the central one, and $r_w=9'$, while the parameters and the definition of assignment of a point to a given area can be modified if necessary. 
     
Thus upon fixing the modes of operation and the CMB temperature threshold range and the step, the geometrical descriptors of the sum, difference, Gaussian or other simulated map can be estimated. All input parameters, auxiliary and temporary data e.g. the numbers of pixels in any given area, and extreme and average values of various parameters can be displayed.

\section{Boomerang maps}

The experiment BOOMERanG - Balloon Observations Of Millimetric Extragalactic Radiation and
Geomagnetism -  measured temperature fluctuations of the Cosmic Microwave Background at
multipole numbers corresponding to the range of the predicted so-called acoustic oscillations at the epoch of the last scattering \cite{bern}. 
The measurements have been performed by means of a millimetric telescope with bolometric detectors located on a balloon borne platform. It was flown in Antarctica in 1998/99 and produced
maps covering 4$\%$ of the sky with high resolution, $\sim 10'$, at four bands from
90 to 410 GHz. 
The rms fluctuation of the anisotropy areas was $\sim 80 \mu K$. The detected
fluctuations were spectrally consistent with the derivative of a
2.735 K blackbody. Masi et al. \cite{masi} had shown that
contamination from local foregrounds was negligible in the maps at
90, 150 and 240 GHz, and that the 410 GHz channel is a good
monitor for dust emission.
The maps have been obtained from the time ordered data using an iterative procedure, 
which properly takes into account the system noise
and produces a maximum likelihood map. Structures 
larger than 10$^o$ have been removed, to
avoid the dominating effects of instrument drifts and 1/f noise.
The map has also been convolved with a Gaussian kernel to obtain a
final FWHM resolution of 22.5 arcmin.

The study of the Boomerang maps at 150 GHz covering around 1\% of the sky have been 
performed by means of the MAP08 code together with P.De Bernardis and the Boomerang team \cite{map}. 

The anisotropy areas have been revealed at each temperature threshold.
MAP08 enabled the analysis of the properties of various subsets of areas, with
given number of pixels, within given interval of pixel numbers, e.g.
from 3 to 200 pixel ones, the distribution of the areas vs number of pixels, etc.
The mean ellipticity of the areas has been estimated via a procedure
of double averaging, first, over the areas at given temperature threshold, then,
over the threshold interval. The dependence of the mean ellipticity
of the anisotropy areas vs the temperature threshold obtained during
that study are shown in Figures 1 and 2.

\begin{figure}[t]
\begin{center}
\epsfxsize=20pc 
\epsfbox{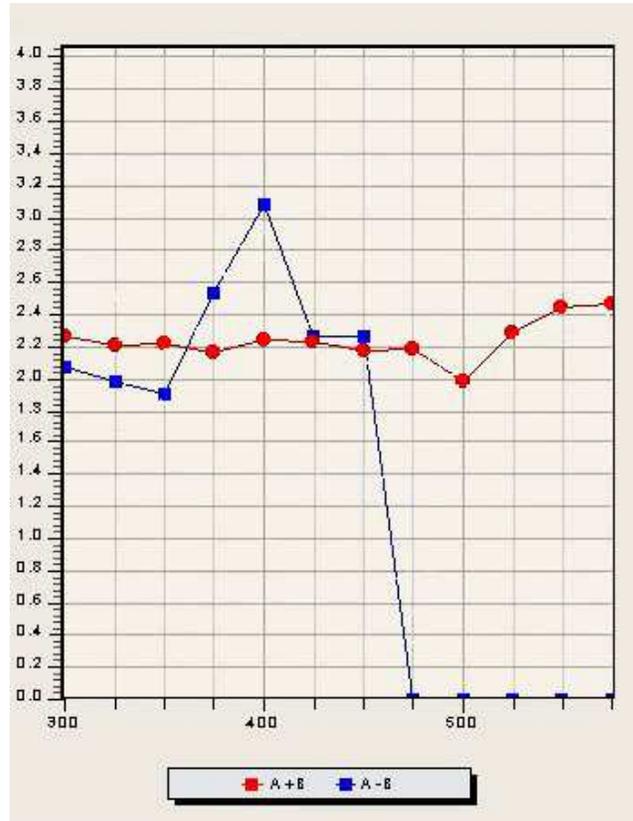} 
\caption{The adaptive software MAP08 revealed the dependence of 
mean ellipticity of anisotropy areas in the Boomerang sum, A+B 
(circles), and difference, A-B (squares), maps created from 
independent channels, on the CMB temperature in $\mu K$; positive 
thresholds. A+B maps contain a cosmological signal, while A-B maps
contain mainly the noise. From 3 to 200-pixel areas are taken into 
account, the step of the matrix MM is 0.6 arc min, (for details and
error boxes see \protect\cite{map}).  
\label{fig:1}}
\end{center}
\end{figure}

\begin{figure}[t]
\begin{center}
\epsfxsize=20pc 
\epsfbox{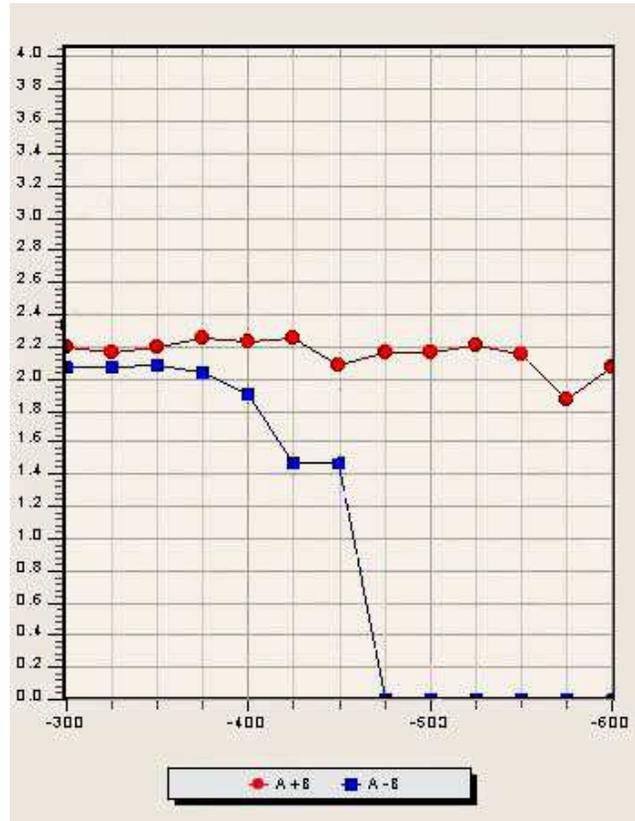} 
\caption{The same as in Fig 1. Negative thresholds.
\label{fig:2}}
\end{center}
\end{figure}

\begin{figure}[t]
\begin{center}
\epsfxsize=20pc 
\epsfbox{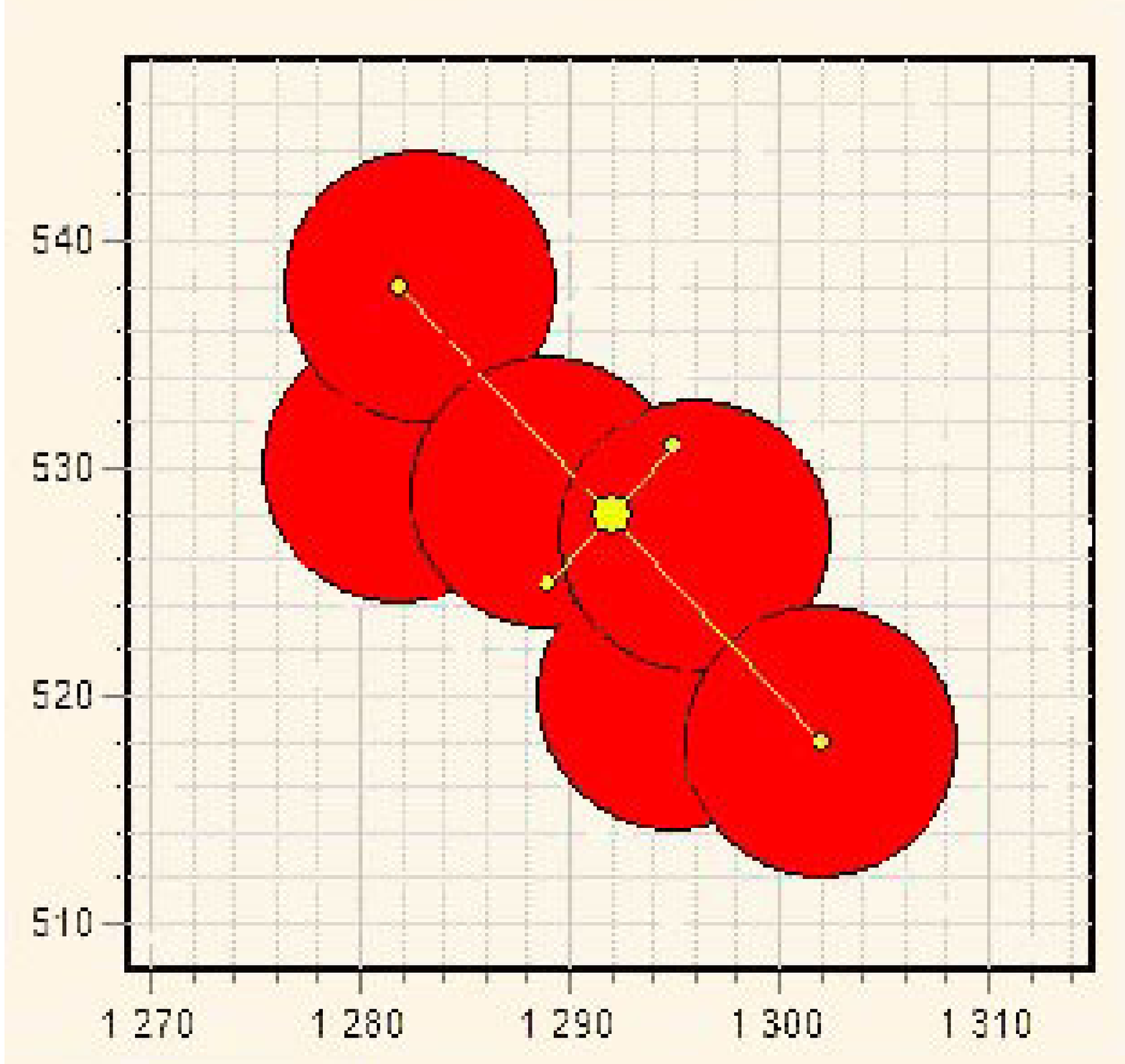} 
\caption{An example of an anisotropy area in Cosmic Microwave 
         Background map observed by Boomerang \protect\cite{bern}. The semi-axes 
         defined for the estimation of the ellipticities
         are shown. 
\label{fig:3}}
\end{center}
\end{figure}

The presence of a threshold interval where the ellipticity is independent on the
threshold is seen from Figure 1. Such behavior was shown to be robust with respect
to the pointing reconstruction procedure accuracy.
The ellipticity due to the noise on the other hand,
has to depend on the threshold. Therefore this indicates correlations
existing in the sky signal. Similar elongation has been detected also for 
COBE-DMR 4-year maps. The mean elongation for COBE data was around 1.9 \cite{gt}, for
Boomerang data it is close to 2.2 \cite{map}.
This is remarkable since COBE-DMR and Boomerang data are quite different
by their angular resolution and noise level.

The described CMB ellipticity (see Figure 3), in principle, can arise due to still unknown processes.
However, if such correlations in the sky signal are due to geodesic mixing, as was predicted, than  this can be interpreted as model-independent indication of two effects, of negative curvature 
and the non-Friedmannian expansion of the Universe \cite{Gur}.

The further work on simulated maps and especially the search of this effect 
at forthcoming more accurate experiments is of particular interest.

\section{Arrows of Time}

Let us mention another aspect of the effect of
geodesic mixing which may provide a condition
necessary for emergence of the thermodynamic arrow of time.
This mechanism can also explain why CMB  contains the
major fraction of the entropy of the Universe.

The thermodynamical arrow for a statistical system can be formulated as 
a consequence of the following conditions (see e.g. \cite{penrose}, \cite{pp}, \cite{zurek},  \cite{zeh}):

1) Decorrelated (special) initial conditions;

2) No memory dynamics.

It should be emphasized that both these conditions are necessary and they
appear already in the Boltzmann's derivation of his kinetic equation, though not explicitly.
They can be traced clearly in Zwanzig's derivation of
master-equation \cite{zwanzig} or Jaynes' information-theoretical
approach to irreversibility \cite{jaynes}. A usual discussion about
possible relations between cosmological and thermodynamical arrows of time
concentrates mainly on the first condition \cite{penrose}, 
\cite{zeh}, \cite{mackay}.
However, one can show \cite{allgur} that this is not sufficient, since 
{\it special initial conditions alone can generate only a thermodynamical pre-arrow of time}.

We then point out that, along with the initial conditions
the second ingredient of the thermodynamical arrow can have a cosmological
context as well, due to mixing of null trajectories in
hyperbolic spaces.

If this is indeed the mechanism of the origin of the thermodynamic arrow,
then the thermodynamics in a flat and positively curved universes not
necessarily to be strongly time asymmetric, and the latter is observed since
we happened to live in a Universe with negative curvature.
In other words, the symmetry of the Newtonian mechanics,
electrodynamics, quantum mechanics might purely survive in some universes.
On the other hand, a recent activity devoted to the foundations of
thermodynamics
allows to disentangle time-asymmetric elements from the remained basis.

In this context the essence of thermodynamical arrow must be understood as
not the mere
increase of entropy of an almost closed system, but the fact that this arrow
has the universal direction in the entire Universe (see \cite{GMH}).
In the light of the suggested explanation
of the emergence of this arrow, it follows that the negative curvature is
the very mechanism unifying all local thermodynamical arrows. While
in the flat or positively curved universes, i.e. at the absence of a 
global unification mechanism, there can be local thermodynamical arrows
with various directions. 

This enabled us to formulate the {\it curvature anthropic principle}, to reflect
the difference of conditions for life in the hyperbolic Universe 
and hence with unified
arrows, and in flat or positively curved spaces, i.e.
at the absence of such unification mechanism \cite{allgur}. 

Often
the thermodynamical arrow of time is identified with the second law of thermodynamics,
and the appearance of Gibbs distribution. We show \cite{allgur} that the 
second law, and the Gibbs distribution can be obtained from 
purely time-symmetric arguments, and need not be consequences of the 
thermodynamical arrow. 

Thus CMB has to carry the direct signature of the thermodynamical and
cosmological arrows.

\section{Conclusion}

We showed that the algorithmic information approach can enable not only qualitative but also
quantitative study of astrophysical problems. The estimation of the
Kolmogorov complexity for 
computer-generated CMB maps and detection of
threshold independent ellipticity in the COBE and Boomerang sum maps, geodesic mixing and
possible model-independent indication of the non-zero negative curvature
and the accelerated expansion of the Universe,
show the efficiency of the approach.

One may expect the further use of algorithmic information concepts not only in
fundamental problems but also in various applications.  
The seeds of such developments are seen already now, considered 
as fiction several decades ago.
For example, instead of sending a letter by post, now
it is enough to send a binary coded signal which then is
transfered to a hardcopy. The same is true for a color image, music, movie. 
In principal the same operation can be
performed, say with an apple, via sending the relevant complete information.
One may think, that in future even human beings can travel via
transfer of information, thus realizing the speed-of-light travels.

\vspace{0.in}

I am grateful to many colleagues. Thus, many of the mentioned results have been obtained
together with my collaborators A.Kocharyan, A.Allahverdyan and A.Kashin.
The Boomerang data have
been analyzed together with P.de Bernardis and the Boomerang team.
Numerous discussions with R.Penrose were of particular importance.

\newpage

\vskip 1cm

{\bf DISCUSSION ON THE REPORT BY V. G. GURZADYAN } \\

{\it Chairman: A. Bohm}

\vskip 5mm

{\bf L. Reichl:}
Do you have a result? Does the universe have negative curvature?

{\bf V. G. Gurzadyan:}
We have seen an effect in cosmic background radiation maps from the analysis of experimental data of COBE and Boomerang without any model dependent approach. It is not a fitting of data by a curve of a given model. The recent models contain 11-13 free parameters describing the content of the matter in the universe, of the dark matter, the initial fluctuation spectrum, etc.
To fit an empirical curve by a model with so many free parameters is not 
as difficult. The problem is to prove that this is the only possible model.
Our approach is model independent.
We can say that we have a new effect, which would exist in non-flat universe.
Therefore either we have a new physical effect or this effect is due to 
non-zero curvature of the universe.

{\bf I. M. Khalatnikoff:} Non zero, negative or positive?

{\bf V. G. Gurzadyan:} Negative.

{\bf I. M. Khalatnikoff:} Closed or open?

{\bf V. G. Gurzadyan:} I prefer not to use the words ``closed'' or ``open'' because Einstein 
equations define the geometry but not the topology of the universe.
For example, at zero curvature one can have a sheet, a cylinder, 
a torus. We can speak only on the negative curvature
$k=-1$, but topologically the universe can be both compact (closed) or non-compact (open).

{\bf L. Stodolsky:} The usual opinion in astrophysical circles is that the curvature is equal to zero.
So, you have found it different?

{\bf V. G. Gurzadyan:} Yes, at present flat models are preferred, in part as inflationary motivated, though there are claims for other
models as well. Precise flatness cannot be proved however, not only due to the observational errors, but due to the degeneracy and dependence on
a number of free parameters.

{\bf M. Courbage:}
Does the complexity depend on the stage of the evolution of the universe?

{\bf V. G. Gurzadyan:}
Of course. I have shown a formula where the complexity depends on the 
time from last scattering epoch.
The effect of geodesic mixing could be observed, in principle, from quasars.
But there is no enough time (distance) for photons from quasars to feel the
geometry. Cosmic background photons are moving too long.
The problem comes to the numerical measurement of the tiny effect, whether it is possible or no. It appears that it can be possible.

{\bf I. M. Khalatnikoff:} 
You have shown us different geometry of spots. 
Have you concluded about the curvature from the analysis of these spots?

{\bf V. G. Gurzadyan:}
The analysis was motivated by the predicted effect and we have 
found its signature. 
It may be a signature of another effect.

{\bf P. Stamp:}
It is a way of calculation of multiple correlations between the densities 
of radiation and matter. This was calculated since 1962.
Can you, from these correlation functions, which you can simply extract
from measurements, deduce the curvature $k$?

{\bf V. G. Gurzadyan:}
If you mean the correlations in the angular power spectrum, the acoustic peaks, they were indeed predicted long time ago, most clearly by Doroshkevich, Sunyaev and Zeldovich in 1978. They are now measured by Boomerang and at other experiments. The autocorrelation function indeed depends on the curvature but also on many other parameters and though provides important constraints on the curvature, the deduction of the precise value of $k$ is not as simple.
Here I discussed correlations in the sky maps.

\end{document}